\documentclass[
  aps,
  prl,
  reprint,            
  superscriptaddress, 
  amsmath,amssymb,   
  floatfix            
]{revtex4-2}

\usepackage{silence}
\usepackage[utf8]{inputenc}
\usepackage{placeins}
\usepackage{amsmath}
\usepackage{amssymb}
\usepackage{graphicx}
\usepackage[normalem]{ulem}
\usepackage{dcolumn}  
\usepackage{bm}       
\usepackage{orcidlink}

\usepackage{hyperref}
\hypersetup{
    colorlinks=true,
    linkcolor=blue,
    citecolor=blue,
    urlcolor=blue
}

\newcommand{\SN}{SN} 
\newcommand{\crossing}{\textit{nexus}}

\begin{document}

\title{Deus ex \texorpdfstring{$H_0$}{H0} -- Is evidence for dynamical dark energy conditioned on early cosmology?
}

\author{Nils Sch\"oneberg\orcidlink{0000-0002-7873-0404}}
\affiliation{University Observatory, Faculty of Physics, Ludwig-Maximilians-Universit\"at,
Scheinerstr. 1, 81677 M\"unchen, Germany}
\affiliation{Excellence Cluster ORIGINS, Boltzmannstrasse 2, 85748 Garching, Germany}

\author{Rodrigo Calder\'on\orcidlink{0000-0002-8215-7292}}\affiliation{CEICO/FZU, Institute of Physics of the Czech Academy of Sciences, Prague, Czech Republic}

\author{Julien Lesgourgues\orcidlink{0000-0001-7627-353X}}
\affiliation{Institute for Theoretical Particle Physics and Cosmology (TTK), RWTH Aachen
University, D-52056 Aachen, Germany}

\date{\today}

\begin{abstract}
Some free-form reconstructions of the dark energy equation of state suggest that dynamical dark energy is the only explanation for the observed data. In this letter we argue that early Universe solutions to the Hubble tension (around or before recombination) generically cause the evidence for this claim to strongly reduce, establishing a tight connection between the early and late cosmology. In particular, the level of evidence for dynamical dark energy depends on the parameters $H_0 r_\mathrm{d}$ and $\Omega_\mathrm{m}$ and early universe solutions typically push towards higher values of $H_0 r_\mathrm{d}$ and lower values of $\Omega_\mathrm{m}$\,, where such evidence is reduced.
\end{abstract}

\hfill \texttt{TTK-26-26}
\maketitle

\enlargethispage*{2\baselineskip}
\emph{Introduction.---} The evidence for dynamical dark energy from model-independent analyses with DESI BAO and SN data (e.g.,~\cite{DESI:2025fii,DESI:2024aqx,DESI:2025wyn,Sharma:2025iux,GuptaChoudhury:2026gsl}) appears overwhelming.
Yet, model-specific analyses like Refs.~\cite{Lynch:2024hzh,Mirpoorian:2025rfp,Chaussidon:2025npr,SPT-3G:2025vyw,Jhaveri:2026bla,Schoneberg:2026vaf} argue that early universe solutions to the Hubble tension (like modified recombination and early dark energy) fit cosmological data typically as well as a dynamical dark energy models. Ref.~\cite{Jhaveri:2026bla} also shows that the combination of early dark energy plus thawing dark energy explains the expansion history data as well a phantom-crossing model, while also reducing the Hubble tension. In this short letter, we argue that these analyses represent different sides of the same coin and are connected through a very generic argument.

\begin{figure*}[t]
    \centering
    \includegraphics[width=0.7\linewidth]{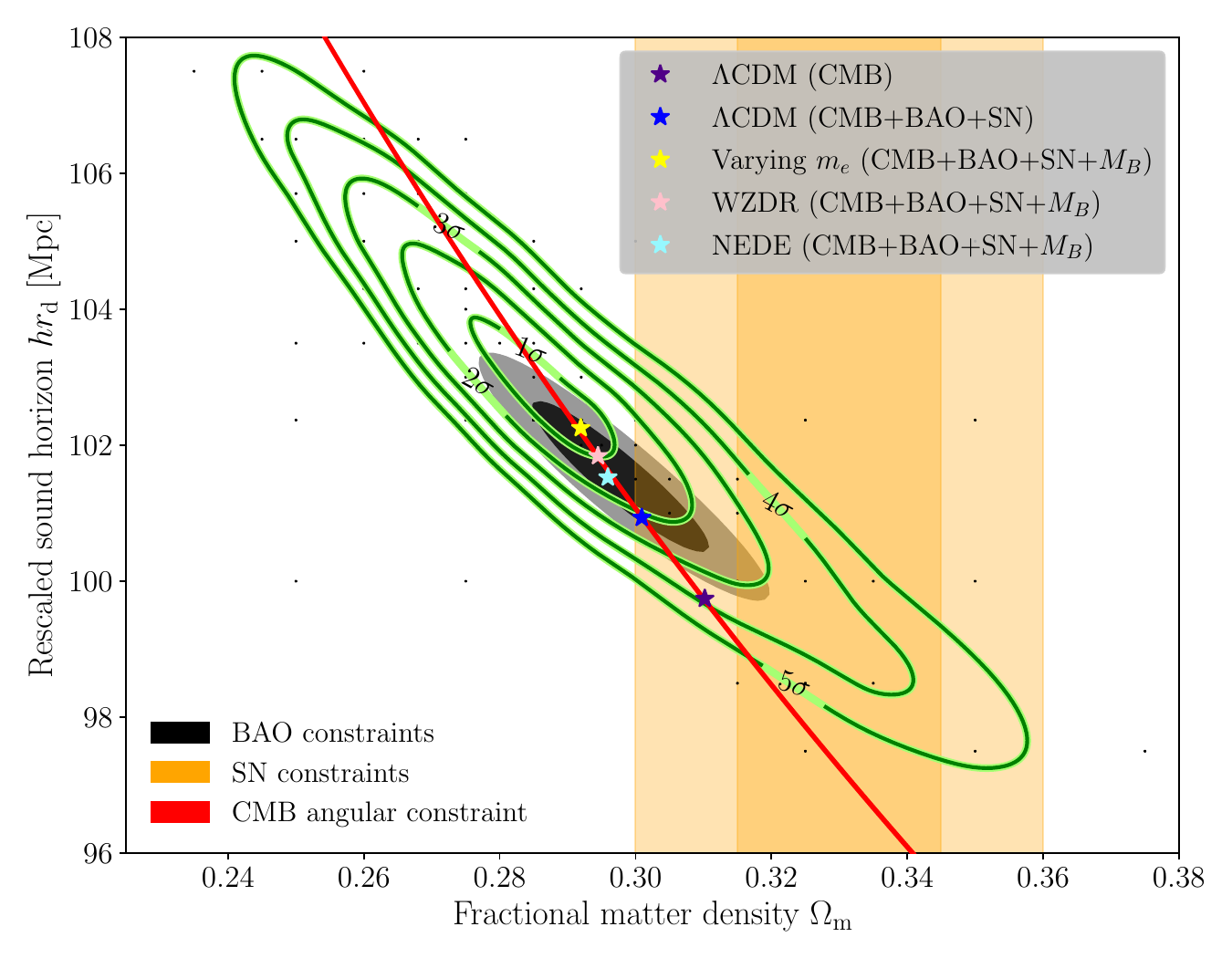}
    \caption{The colored regions show the 68\% and 95\% credible intervals for $(\Omega_\mathrm{m}, h r_\mathrm{d})$ inferred from DESI DR2 BAO (black) or DES Y5 Dovekie uncalibrated SNIa (orange) when assuming a late-time $\Lambda$CDM expansion history. The thin red band shows the 68\% constraint from the CMB peak scale under the same assumption (this band is nearly independent of  early universe assumptions, up to the small caveats explained in the text). The green iso-contours show the level of preference for CPL dark energy against a plain cosmological constant given BAO and SN data and for given values of $(\Omega_\mathrm{m}, h r_\mathrm{d})$ (see the supplementary material for details). The stars show the best-fit values when fitting $\Lambda$CDM and selected extended models to CMB+BAO+\SN{} data (including also $M_B$ measurements for models proposed as possible solutions to the Hubble tension) taken from Ref.~\cite{Schoneberg:2026vaf}.}
    \label{fig:rdh_desdovekie}
\end{figure*}

\emph{Geometrical constraints from BAO and \SN{}.---} Measurements of the BAO at redhsift $z_i$ constrain both the angular ($\Delta \theta_i$) and the line-of-sight extent ($\Delta z_i$) of the acoustic scale. By decomposing the Hubble rate as $H(z) = H_0 E(z)$, it is straightforward to see that only the product $H_0 r_\mathrm{d}$ involving the sound horizon at baryon drag $r_\mathrm{d}$ can be determined from these, in addition to constraints on the expansion rate $E(z)$ through
\begin{align}
    \theta_i = \beta\: (H_0 r_\mathrm{d}) / f(z_i)\ , & & \Delta z_i = \beta\: (H_0 r_\mathrm{d}) \: E(z_i)\ ,
\end{align}
with $f(z) = \int_0^z \mathrm{d}z'/E(z')$ in a flat universe. Here $\beta$ is an additional factor accounting for a phase shift from neutrinos, free-streaming dark radiation, or from the difference between the $r_\mathrm{d}$ integral and template fit outlined in Ref.~\cite{Asensio-Rivera:2026mjd}. 
It is typically very close to one and changing by $\sim 0.1\%$ only for reasonable variations of cosmological parameters (e.g. $\Delta N_\mathrm{eff}<1$) \cite{Asensio-Rivera:2026mjd}.

Similarly, the uncalibrated measurements of \SN{} brightness as a function of redshift measure $(1+z)^2/f(z)$ with only minimal assumptions of general relativity. Any dependence on $H_0$ is completely degenerate with the absolute calibration.

In a $\Lambda$CDM universe, $E(z)$ and thus $f(z)$ only depend on $\Omega_\mathrm{m}$ at low redshift, almost independently of $\sum m_\nu$, since neutrino mass eigenstates with a non-negligible mass become non-relativistic at a redshift higher than the highest $z_i$\,. In extended late-cosmology models, $E(z)$ may also depend on spatial curvature ($\Omega_\mathrm{k}$), dynamical dark energy parameters (e.g., $w_0$, $w_a$), or possibly some non-standard dark matter decay rate. 

For an expansion history with a cosmological constant and CDM, the BAO and \SN{} constraints can be entirely summarized in ($\Omega_\mathrm{m}$, $h r_\mathrm{d}$) space, $h$ being the reduced Hubble constant $h = H_0/\mathrm{100km/s/Mpc}$. They are shown as black and orange contours in Figure~\ref{fig:rdh_desdovekie}. Notably, individual BAO measurements provide allowed regions with different slopes between $h r_\mathrm{d}$ and $\Omega_\mathrm{m}$\,, but their intersection reduces to a nearly elliptical preferred region \cite{DESI:2025zgx}. There is a mild $\sim 1.8\sigma$ tension between the BAO and \SN{} contours with this expansion rate, as highlighted also in Ref.~\cite{Shlivko:2026jxa}. As argued above, this tension can be mitigated only by late universe physics as it only depends on $E(z)$ for redshifts $z \lesssim 2.4$.

\emph{Constraints from the CMB.---}
Full-shape CMB constraints yield information that extends beyond mere geometric distance measures. However, many early-universe models introduce a surprising flexibility in final CMB parameter constraints \cite{Schoneberg:2021qvd,Khalife:2023qbu,Schoneberg:2026vaf}, raising the question of whether any constraint can be extracted model-independently. Given that the CMB angular scale is directly measured by the peak positions in the angular power spectrum, though, it remains exceptionally robust to model assumptions. It measures
\begin{equation}
    \theta_\mathrm{CMB} = \beta_\mathrm{drag} \beta\: (H_0 r_\mathrm{d}) /f(z_\mathrm{CMB})\ .
\end{equation}
Here $\beta_\mathrm{drag}$ is the ratio of sound horizon scales at photon decoupling and baryon drag, $\beta_\mathrm{drag} = r_\mathrm{s}/r_\mathrm{d}$\,. Unlike $f(z_i)$ for BAO, $f(z_\mathrm{CMB})$ depends also on effects in the early Universe parametrized by the summed neutrino mass, the matter-radiation equality, and the precise recombination redshift. In practice, these additional dependencies have a remarkably small impact, with $\beta_\mathrm{drag}$ and $f(z_\mathrm{CMB})$ both varying for a fixed value of $\Omega_\mathrm{m}$ by less than 0.1\% for $\Delta \sum m_\nu = 1\mathrm{eV}$ or $\Delta N_\mathrm{eff}=1$, and the latter by less than 0.5\% for $\Delta z_\mathrm{CMB}=5$. 
Given the incredibly tight constraints from the CMB (making the $1\sigma$ constraints in Figure~\ref{fig:rdh_desdovekie} look like a thin line), these tiny variations may be relevant for CMB constraints on several cosmological parameters. However, they only have a very minor impact on the discussions below, based on comparatively much looser BAO+SN constraints.

\enlargethispage*{2\baselineskip}
Assuming an expansion rate with a cosmological constant and CDM, it is remarkable that the CMB angular constraint crosses right through the middle of the BAO constraint. Yet, moving along the CMB degeneracy line (by employing different early Universe models), it is impossible to reach simultaneous perfect agreement with both the BAO and the \SN{} region. 
Assuming standard $\Lambda$CDM, the combination of Planck, SPT, and ACT data (CMB-SPA) gives a best fit at
$h r_\mathrm{d} = (99.7 \pm 1.1)$Mpc and $\Omega_\mathrm{m} = 0.3102 \pm 0.0087$ (centered on the white star in Figure~\ref{fig:rdh_desdovekie}), 
in slight tension with the BAO contours. 
The combination of CMB with BAO and SN data prefers a compromise region centered on $h r_\mathrm{d} = (100.93\pm0.46)$Mpc, $\Omega_\mathrm{m}=0.3009\pm0.0034$ (centered on the blue star).

Instead, many models with a different early evolution -- often discussed in the Hubble tension context -- prefer larger values of $h r_\mathrm{d}$ and smaller values of $\Omega_\mathrm{m}$. This is a natural consequence of increasing $h$ while not increasing $\Omega_\mathrm{m} h^2$ enough to compensate, leading to smaller values of $\Omega_\mathrm{m}$\,, which forces higher $h r_\mathrm{d}$ through the geometric constraints \cite{Poulin:2024ken}.

We can now pose the question of how strongly a deviation from the cosmological constant is preferred given the BAO+SN data set for a given $(h r_\mathrm{d}, \Omega_\mathrm{m})$ and independently of early universe models. We do this using the Chevallier-Polarski-Linder (CPL) \cite{Chevallier:2000qy,Linder:2002et} parameterization of the dark energy equation of state $w(a) = w_0 + w_a (1-a)$ for scale factor $a=1/(1+z)$ and measuring the Mahalanobis distance \cite{mahalanobis1936generalized} of the BAO+SN contour from $w_0=-1, w_a=0$, which we convert into a significance measured in numbers of sigma (see the supplementary material for more details).  We show the iso-contours of this significance of departure from a cosmological constant as green contour lines in Figure~\ref{fig:rdh_desdovekie}. Interestingly, the region where this preference is minimized -- below $0.5\sigma$ -- is exactly where extended early models are moving. We dub this region the \crossing{}. 
It is centered on $h r_\mathrm{d} \approx 103\mathrm{Mpc}$ and $\Omega_\mathrm{m} \approx 0.285$ for the DES Y5 Dovekie \cite{DES:2025sig} dataset. Equivalent results for Pantheon+ \cite{Brout:2022vxf} and Union3 \cite{Rubin:2023jdq} are shown in Appendix~\ref{supplementary}.

This region is inferred from a global BAO+SN fit, but appears visually as a better fit to BAO than SN data alone, due to the stronger constraining power of the former dataset. Additionally, the fact that the \crossing{} is centered still within the BAO contours shows that there is as of yet little in the shape of BAO or BAO+SN measurements (as a function of $z$) that prefers dynamical dark energy: a strong preference arises instead when assuming a value of $h r_\mathrm{d}$ departing from the \crossing{} region.

\begin{figure}
    \centering
    \includegraphics[width=1\linewidth]{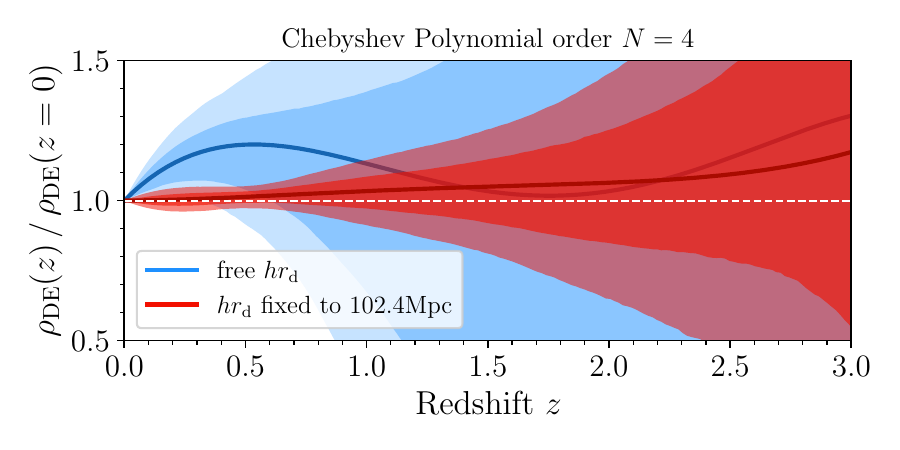}
    \caption{Reconstruction of the dark energy density based on a fourth-order Chebyshev expansion (as in \cite{DESI:2024aqx,DESI:2025fii}), comparing between a model in which $h r_\mathrm{d}$ and $\Omega_\mathrm{m}$ are allowed to take on any value (free, light blue), and one where we fix $h r_\mathrm{d}$ and $\Omega_\mathrm{m}$ to the value of the yellow star in Figure~\ref{fig:rdh_desdovekie} (fixed, red). Based on BAO + Union3 SN data.}
    \label{fig:N4}
\end{figure}

This leads us to the important conclusion that the evidence for dynamical dark energy is conditioned on assumptions about the early universe cosmology.
We confirm this with a full reconstruction of the dark energy density based on a fourth-order Chebyshev polynomial expansion, presented in Figure~\ref{fig:N4} (see \cite{DESI:2024aqx} for further details). The darker (red) contours show that the evidence for dynamical dark energy weakens considerably close to the \crossing. In this case we use the value of $(\Omega_\mathrm{m}, h r_\mathrm{d})$ for the particular model shown as a yellow star in Figure.~\ref{fig:rdh_desdovekie}.

This might seem counterintuitive: At first sight, the analyses of Refs.~\cite{DESI:2025fii,Sharma:2025iux,GuptaChoudhury:2026gsl} seem to
suggest otherwise. However, Ref.~\cite{GuptaChoudhury:2026gsl} uses a $\Lambda$CDM prior on $r_\mathrm{d}$, while in Ref.~\cite{DESI:2025fii} $r_\mathrm{d}$ is determined based on its $\Lambda$CDM expression at early times with priors on $\Omega_\mathrm{b} h^2$ and $\Omega_\mathrm{cdm} h^2$. In both analyses, early universe cosmologies significantly shifting $r_\mathrm{d}$ are not represented. In Ref.~\cite{Sharma:2025iux}, the results are independent of the sound horizon, but not of the assumed cosmology. A subtle interplay of the priors on $A_\mathrm{s}$ and $n_\mathrm{s}$ adopted in Ref.~\cite{Sharma:2025iux} with CMB lensing and galaxy lensing data leads to an implicit constraint on $\Omega_\mathrm{m}$ that keeps the models away from the \crossing{} and these priors do not hold for many early cosmology models.

In the analysis of Ref.~\cite{Schoneberg:2026vaf}, a $1.3\sigma$ preference for dynamical dark energy is pointed out, which is larger than what Figure.~\ref{fig:rdh_desdovekie} suggests at the yellow star -- this is because the contours in Ref.~\cite{Schoneberg:2026vaf} are centered at $h r_\mathrm{d} = (101.17 \pm 0.81)\mathrm{Mpc}$ and $\Omega_\mathrm{m}=0.2986 \pm 0.0053$ (much lower than the yellow star) once allowing for CPL-parameterized dark energy. We explain this apparent contradiction below.

\emph{Supernovae strike back.---} It is important to stress that the statement of the existence of the \crossing{} is not a silver bullet; the \crossing{} sits at lower $\Omega_\mathrm{m}$ and is thus in tension with \SN{} data, leading to a residual tension between CMB+BAO and \SN{} data. Early universe models thus only converge towards the region of parameter space of increased $h r_\mathrm{d}$ and decreased $\Omega_\mathrm{m}$ when they are forced to reduce the Hubble tension. Then, they generically also reduce the significance of the preference for dynamical dark energy (see also \cite[Tab.~VII]{Schoneberg:2026vaf}).

\enlargethispage*{1\baselineskip}
Overall, roughly speaking, one trades a reduction of $\sim 2.5\sigma$ to $1.3\sigma$ preference for dynamical dark energy (e.g. from the blue to yellow points in Figure~\ref{fig:rdh_desdovekie}) for a raise from $1.8\sigma$ to $2.3\sigma$ tension in $\Omega_\mathrm{m}$ with respect to SN data -- but at the same time, this can potentially lead to a reduction of the Hubble tension from $\sim 4.2\sigma$ to $\sim 1.9\sigma$. In terms of $\Delta \chi^2$, this can be a somewhat viable trade, but it is not necessarily entirely satisfying.

\begin{figure}
    \centering
    \includegraphics[width=0.8\linewidth]{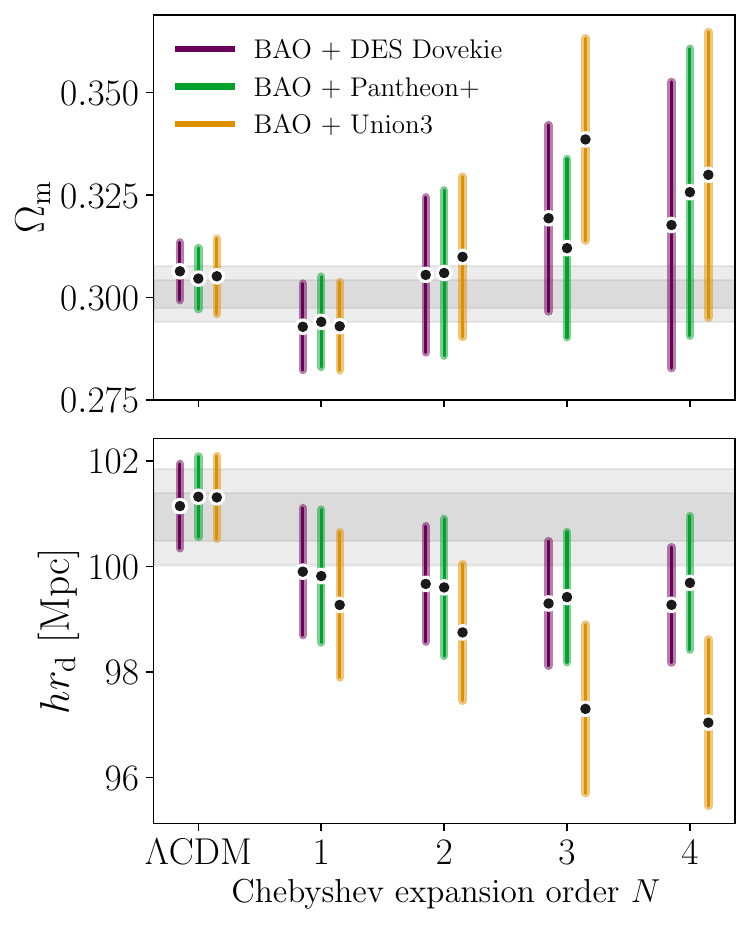}
    \caption{Mean and standard deviation of $\Omega_\mathrm{m}$ and $h r_\mathrm{d}$ when fitting only BAO and SN data, for a model of $E(z)$ with cold dark matter and various orders of a Chebyshev polynomial expansion of the dark energy density (normalized to its present value). The grey contours show the 68\% and 95\% constraints on $h r_\mathrm{d}$ and $\Omega_\mathrm{m}$ from the combination of CMB+BAO+SN data in the $\Lambda$CDM model for reference. They are taken from Ref.~\cite{Schoneberg:2026vaf}.}
    \label{fig:late_only}
\end{figure}

In particular, if we only allowed the BAO and uncalibrated SN data to guide our investigations, we would conclude that dynamical dark energy is preferred, together with lower $h r_\mathrm{d}$ and higher $\Omega_\mathrm{m}$ values than at the \crossing. We show this in Figure~\ref{fig:late_only} for a Chebyshev expansion (as in Ref.~\cite{DESI:2024aqx}) of the dark energy density. However, if we accept the Hubble tension as evidence for a mechanism requiring a lower $\Omega_\mathrm{m}$ and higher $h r_\mathrm{d}$\,, we are necessarily pushed towards the \crossing{}, with a value of $\Omega_\mathrm{m}$ more in tension with supernovae, and weaker evidence for dynamical dark energy irrespective of our preference.

\emph{Conclusions.---} 
Overall, it is worth highlighting that the preference for dynamical dark energy is not a model-independent statement, and is instead conditioned on whatever early universe physics is assumed. Models that solve the Hubble tension necessarily push the investigation to trade a $>2\sigma$ preference for DDE for an $0.5\sigma$ increase in the SN tension. Therefore, statements about the universe necessarily involving a phantom crossing or even thawing dark energy are a little premature, and new late universe data will be required to pin this down.

There will be further opportunities to measure $\Omega_\mathrm{m}$ independently and determine on which point of the CMB geometrical constraint the cosmological model is sitting. The measurement of the full shape of the galaxy spectrum will be one of them. For instance, ShapeFit \cite{Brieden:2021edu,Lai:2024bpl} will return a nearly model-independent measurement of $\Omega_\mathrm{m}$ (although relying on an $n_\mathrm{s}$ prior as well). The measurement of the Lyman-$\alpha$ flux spectrum will also be sensitive to $\Omega_\mathrm{m}$ through $S_8$.

Finally, future SN data and BAO measurements will bring crucial constraints on the late expansion history, and reveal whether the slight $\Omega_\mathrm{m}$ tension seen in Figure~\ref{fig:rdh_desdovekie} persists or not. If the true cosmological model departs from $\Lambda$CDM at early times and has a high value of $h r_\mathrm{d}$, SN data will eventually favor lower values of $\Omega_\mathrm{m}$, as in the days of Pantheon \cite{Pan-STARRS1:2017jku}. Otherwise, a whole class of solutions to the Hubble tension might be severely restricted.
\FloatBarrier


\section*{Acknowledgements}
This work emerged from interactions at the workshop ``Exploring New Frontiers in Cosmology'' at the GGI Institute, Florence. We thank the Institute and workshop organizers for their hospitality. We also thank Eric Linder and Vivian Poulin for very helpful comments on the first version of this draft. The authors acknowledge the use Google's Gemini 3.6 for aiding in plotting and data analysis. We acknowledge the use of the Python packages \texttt{liquidcosmo}, \texttt{classy}, and \texttt{cobaya}.
NS acknowledges support from the Excellence Cluster ORIGINS which is funded by the Deutsche Forschungsgemeinschaft (DFG, German Research Foundation) under Germany’s Excellence Strategy - EXC-2094/2 - 390783311, as well as the funding through a Fraunhofer-Schwarzschild Fellowship at the LMU. R.C. is funded by the Czech Ministry of Education, Youth and Sports (MEYS) and European Structural and Investment Funds (ESIF) under project number CZ.02.01.01/00/22\_008/0004632.

\bibliography{biblio} 

@article{Asensio-Rivera:2026mjd,
    author = {Asensio-Rivera, Francisco and Sch{\"o}neberg, Nils and Gil-Mar{\'\i}n, H{\'e}ctor and Verde, Licia},
    title = "{The BAO scale {\textemdash} how standard is the standard ruler?}",
    eprint = "2603.03443",
    archivePrefix = "arXiv",
    primaryClass = "astro-ph.CO",
    doi = "10.1088/1475-7516/2026/07/046",
    journal = "JCAP",
    volume = "2026",
    number = "07",
    pages = "046",
    year = "2026"
}

@article{Shlivko:2026jxa,
    author = "Shlivko, David and Poulin, Vivian",
    title = "{Phantom-Crossing Dark Energy and the $\Omega_m$ Tug-of-War}",
    eprint = "2603.22406",
    journal = "arXiv e-prints",
    archivePrefix = "arXiv",
    primaryClass = "astro-ph.CO",
    month = "3",
    year = "2026"
}

@article{GuptaChoudhury:2026gsl,
    author = "Gupta Choudhury, Shibendu and Mukherjee, Purba and Di Valentino, Eleonora and Sen, Anjan A.",
    title = "{Model-Independent Indication for a Localized Anomaly in the Late-Time Expansion History}",
    eprint = "2607.13009",
    archivePrefix = "arXiv",
    primaryClass = "astro-ph.CO",
    journal = "arXiv e-prints",
    month = "7",
    year = "2026"
}

@article{Rubin:2023jdq,
    author = "Rubin, David and others",
    title = "{Union Through UNITY: Cosmology with 2,000 SNe Using a Unified Bayesian Framework}",
    eprint = "2311.12098",
    archivePrefix = "arXiv",
    primaryClass = "astro-ph.CO",
    doi = "10.3847/1538-4357/adc0a5",
    journal = "Astrophys. J.",
    volume = "986",
    number = "2",
    pages = "231",
    year = "2025"
}

@article{Pan-STARRS1:2017jku,
    author = "Scolnic, D. M. and others",
    collaboration = "Pan-STARRS1",
    title = "{The Complete Light-curve Sample of Spectroscopically Confirmed SNe Ia from Pan-STARRS1 and Cosmological Constraints from the Combined Pantheon Sample}",
    eprint = "1710.00845",
    archivePrefix = "arXiv",
    primaryClass = "astro-ph.CO",
    doi = "10.3847/1538-4357/aab9bb",
    journal = "Astrophys. J.",
    volume = "859",
    number = "2",
    pages = "101",
    year = "2018"
}

@article{Lai:2024bpl,
    author = "Lai, Y. and others",
    title = "{A comparison between ShapeFit compression and Full-Modelling method with PyBird for DESI 2024 and beyond}",
    eprint = "2404.07283",
    archivePrefix = "arXiv",
    primaryClass = "astro-ph.CO",
    doi = "10.1088/1475-7516/2025/01/139",
    journal = "JCAP",
    volume = "2025",
    number = "01",
    pages = "139",
    year = "2025"
}

@article{Brieden:2021edu,
    author = "Brieden, Samuel and Gil-Mar{\'\i}n, H{\'e}ctor and Verde, Licia",
    title = "{ShapeFit: extracting the power spectrum shape information in galaxy surveys beyond BAO and RSD}",
    eprint = "2106.07641",
    archivePrefix = "arXiv",
    primaryClass = "astro-ph.CO",
    doi = "10.1088/1475-7516/2021/12/054",
    journal = "JCAP",
    volume = "12",
    number = "12",
    pages = "054",
    year = "2021"
}

@article{DES:2025sig,
    author = "Popovic, B. and others",
    collaboration = "DES",
    title = "{The Dark Energy Survey supernova program: a reanalysis of cosmology results and evidence for evolving dark energy with an updated Type Ia supernova calibration}",
    eprint = "2511.07517",
    archivePrefix = "arXiv",
    primaryClass = "astro-ph.CO",
    reportNumber = "FERMILAB-PUB-25-0842-CSAID-PPD",
    doi = "10.1093/mnras/stag632",
    journal = "Mon. Not. Roy. Astron. Soc.",
    volume = "548",
    number = "4",
    pages = "stag632",
    year = "2026"
}

@article{Jhaveri:2026bla,
    author = "Jhaveri, Tanisha and Karwal, Tanvi and Crawford, Thomas and Hu, Wayne and Khalife, Ali Rida and Balkenhol, Lennart and Ge, Fei",
    title = "{Disentangling cosmic distance tensions with early and late dark energy}",
    eprint = "2604.08530",
    archivePrefix = "arXiv",
    primaryClass = "astro-ph.CO",
    journal = "arXiv e-prints",
    month = "4",
    year = "2026"
}

@article{Mirpoorian:2025rfp,
    author = "Mirpoorian, Seyed Hamidreza and Jedamzik, Karsten and Pogosian, Levon",
    title = "{Is dynamical dark energy necessary? DESI BAO and modified recombination}",
    eprint = "2504.15274",
    archivePrefix = "arXiv",
    primaryClass = "astro-ph.CO",
    doi = "10.1088/1475-7516/2025/12/050",
    journal = "JCAP",
    volume="2025",
    number= "12",
    pages = "050",
    year = "2025"
}

@article{SPT-3G:2025vyw,
    author = "Khalife, A. R. and others",
    collaboration = "SPT-3G",
    title = "{SPT-3G D1: Axion early dark energy with CMB experiments and DESI observations}",
    eprint = "2507.23355",
    archivePrefix = "arXiv",
    primaryClass = "astro-ph.CO",
    reportNumber = "FERMILAB-PUB-25-0610-PPD",
    doi = "10.1103/8jjr-7hpb",
    journal = "Phys. Rev. D",
    volume = "113",
    number = "10",
    pages = "103546",
    year = "2026"
}

@article{Chaussidon:2025npr,
    author = "Chaussidon, E. and others",
    title = "{Early time solution as an alternative to the late time evolving dark energy with DESI DR2 BAO}",
    eprint = "2503.24343",
    archivePrefix = "arXiv",
    primaryClass = "astro-ph.CO",
    reportNumber = "FERMILAB-PUB-25-0241-PPD",
    doi = "10.1103/xtql-wh3h",
    journal = "Phys. Rev. D",
    volume = "112",
    number = "6",
    pages = "063548",
    year = "2025"
}

@article{Lynch:2024hzh,
    author = "Lynch, Gabriel P. and Knox, Lloyd and Chluba, Jens",
    title = "{DESI observations and the Hubble tension in light of modified recombination}",
    eprint = "2406.10202",
    archivePrefix = "arXiv",
    primaryClass = "astro-ph.CO",
    doi = "10.1103/PhysRevD.110.083538",
    journal = "Phys. Rev. D",
    volume = "110",
    number = "8",
    pages = "083538",
    year = "2024"
}

@article{Brout:2022vxf,
    author = "Brout, Dillon and others",
    title = "{The Pantheon+ Analysis: Cosmological Constraints}",
    eprint = "2202.04077",
    archivePrefix = "arXiv",
    primaryClass = "astro-ph.CO",
    doi = "10.3847/1538-4357/ac8e04",
    journal = "Astrophys. J.",
    volume = "938",
    number = "2",
    pages = "110",
    year = "2022"
}

@article{Sharma:2025iux,
    author = "Sharma, Ravi Kumar and Lesgourgues, Julien",
    title = "{Constraints on neutrino mass and dark energy agnostic to the sound horizon}",
    eprint = "2510.15835",
    archivePrefix = "arXiv",
    primaryClass = "astro-ph.CO",
    reportNumber = "TTK-25-31",
    doi = "10.1088/1475-7516/2026/02/034",
    journal = "JCAP",
    volume = "2026",
    number = "02",
    pages = "034",
    year = "2026"
}

@article{DESI:2025fii,
    author = "Lodha, K. and others",
    collaboration = "DESI",
    title = "{Extended dark energy analysis using DESI DR2 BAO measurements}",
    eprint = "2503.14743",
    archivePrefix = "arXiv",
    primaryClass = "astro-ph.CO",
    reportNumber = "FERMILAB-PUB-25-0164-PPD",
    doi = "10.1103/w4c6-1r5j",
    journal = "Phys. Rev. D",
    volume = "112",
    number = "8",
    pages = "083511",
    year = "2025"
}

@article{Schoneberg:2021qvd,
    author = {Sch{\"o}neberg, Nils and Franco Abell{\'a}n, Guillermo and P{\'e}rez S{\'a}nchez, Andrea and Witte, Samuel J. and Poulin, Vivian and Lesgourgues, Julien},
    title = "{The H0 Olympics: A fair ranking of proposed models}",
    eprint = "2107.10291",
    archivePrefix = "arXiv",
    primaryClass = "astro-ph.CO",
    doi = "10.1016/j.physrep.2022.07.001",
    journal = "Phys. Rept.",
    volume = "984",
    pages = "1--55",
    year = "2022"
}

@article{Khalife:2023qbu,
    author = {Khalife, Ali Rida and Zanjani, Maryam Bahrami and Galli, Silvia and G{\"u}nther, Sven and Lesgourgues, Julien and Benabed, Karim},
    title = "{Review of Hubble tension solutions with new SH0ES and SPT-3G data}",
    eprint = "2312.09814",
    archivePrefix = "arXiv",
    primaryClass = "astro-ph.CO",
    reportNumber = "TTK-23-36",
    doi = "10.1088/1475-7516/2024/04/059",
    journal = "JCAP",
    volume = "2024",
    number = "04",
    pages = "059",
    year = "2024"
}

@article{Schoneberg:2026vaf,
    author = {Sch{\"o}neberg, Nils and Poulin, Vivian and Ferrari, Angelo G. and Finelli, Fabio and Lesgourgues, Julien and Morelli, Luca and Mosbech, Markus R. and Sharma, Ravi Kumar and Simon, Th{\'e}o},
    title = "{The $H_0$ world cup. II. A comprehensive competition between proposed Hubble tension solutions}",
    eprint = "2607.13283",
    archivePrefix = "arXiv",
    primaryClass = "astro-ph.CO",
    reportNumber = "TTK-26-23, TTP26-28",
    month = "7",
    year = "2026",
    journal = "arXiv e-prints",
}

@article{mahalanobis1936generalized,
  title={On the generalized distance in statistics},
  author={Mahalanobis, Prasanta Chandra},
  journal={Proceedings of the National Institute of Sciences of India},
  volume={2},
  number={1},
  pages={49--55},
  year={1936}
}

@article{DESI:2025zgx,
    author = "Abdul Karim, M. and others",
    collaboration = "DESI",
    title = "{DESI DR2 results. II. Measurements of baryon acoustic oscillations and cosmological constraints}",
    eprint = "2503.14738",
    archivePrefix = "arXiv",
    primaryClass = "astro-ph.CO",
    reportNumber = "FERMILAB-PUB-25-0169-PPD",
    doi = "10.1103/tr6y-kpc6",
    journal = "Phys. Rev. D",
    volume = "112",
    number = "8",
    pages = "083515",
    year = "2025"
}

@article{DESI:2024aqx,
    author = "Calderon, R. and others",
    collaboration = "DESI",
    title = "{DESI 2024: reconstructing dark energy using crossing statistics with DESI DR1 BAO data}",
    eprint = "2405.04216",
    archivePrefix = "arXiv",
    primaryClass = "astro-ph.CO",
    doi = "10.1088/1475-7516/2024/10/048",
    journal = "JCAP",
    volume = "2024",
    number = "10",
    pages = "048",
    year = "2024"
}

@article{Poulin:2024ken,
    author = "Poulin, Vivian and Smith, Tristan L. and Calder{\'o}n, Rodrigo and Simon, Th{\'e}o",
    title = "{Implications of the cosmic calibration tension beyond H0 and the synergy between early- and late-time new physics}",
    eprint = "2407.18292",
    archivePrefix = "arXiv",
    primaryClass = "astro-ph.CO",
    doi = "10.1103/PhysRevD.111.083552",
    journal = "Phys. Rev. D",
    volume = "111",
    number = "8",
    pages = "083552",
    year = "2025"
}

@article{Chevallier:2000qy,
    author = "Chevallier, Michel and Polarski, David",
    title = "{Accelerating universes with scaling dark matter}",
    eprint = "gr-qc/0009008",
    archivePrefix = "arXiv",
    doi = "10.1142/S0218271801000822",
    journal = "Int. J. Mod. Phys. D",
    volume = "10",
    pages = "213--224",
    year = "2001"
}

@article{Linder:2002et,
    author = "Linder, Eric V.",
    title = "{Exploring the expansion history of the universe}",
    eprint = "astro-ph/0208512",
    archivePrefix = "arXiv",
    doi = "10.1103/PhysRevLett.90.091301",
    journal = "Phys. Rev. Lett.",
    volume = "90",
    pages = "091301",
    year = "2003"
}

@article{DESI:2025wyn,
    author = "Gu, Gan and others",
    collaboration = "DESI",
    title = "{Dynamical dark energy in light of the DESI DR2 baryonic acoustic oscillations measurements}",
    eprint = "2504.06118",
    archivePrefix = "arXiv",
    primaryClass = "astro-ph.CO",
    reportNumber = "FERMILAB-PUB-25-0235-PPD",
    doi = "10.1038/s41550-025-02669-6",
    journal = "Nature Astron.",
    volume = "9",
    number = "12",
    pages = "1879--1889",
    year = "2025",
    note = "[Erratum: Nature Astron. 9, 1898--1898 (2025)]"
}

\clearpage
\appendix
\section{Supplementary material}
\refstepcounter{section}
\label{supplementary}
\FloatBarrier

\begin{figure}[ht]
    \centering
    \includegraphics[width=1\linewidth]{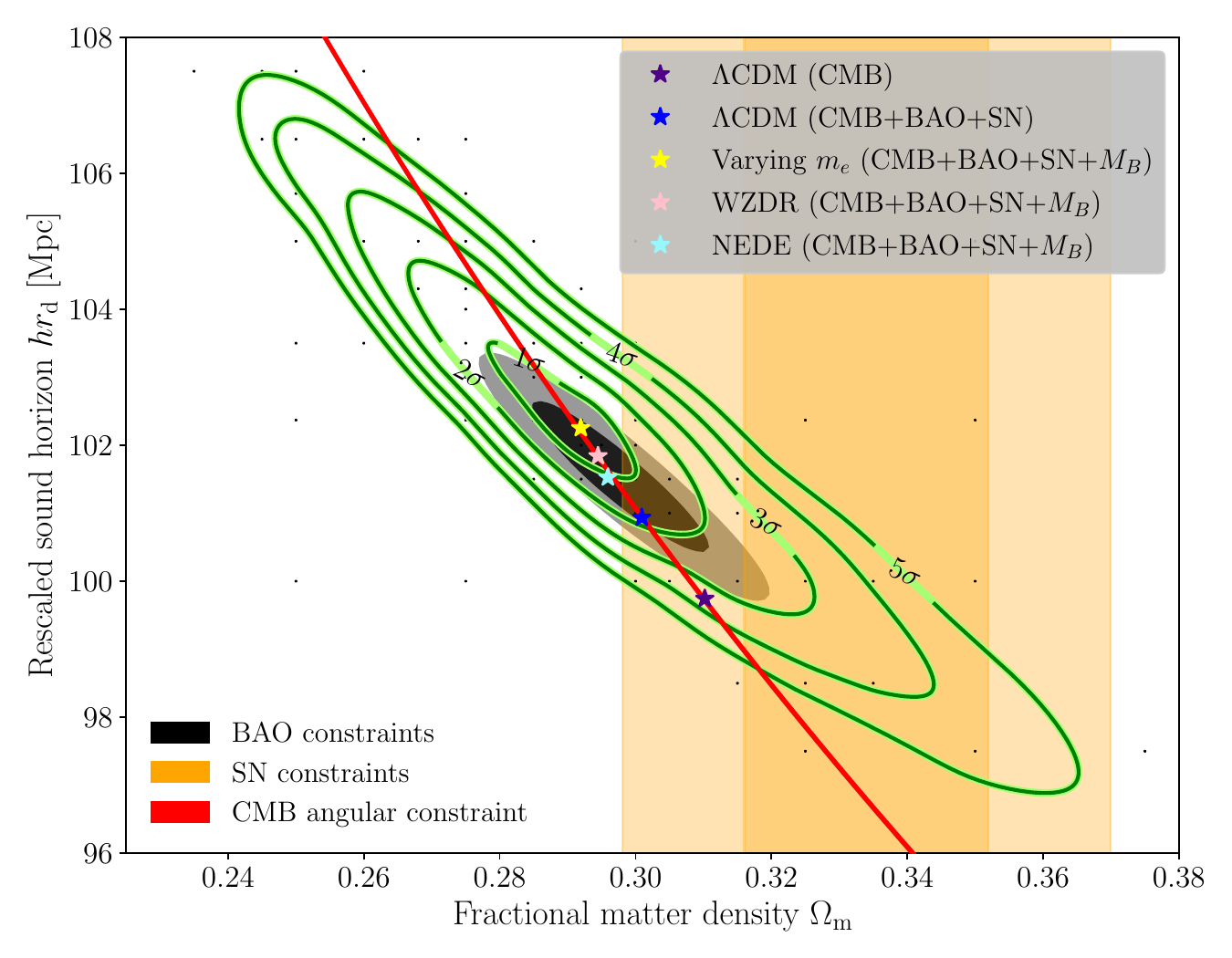}
    \caption{Same as Figure~\ref{fig:rdh_desdovekie} but using Pantheon+ supernovae instead.}
    \label{fig:rdh_pantheonplus}
\end{figure}

\begin{figure}[ht]
    \centering
    \includegraphics[width=1\linewidth]{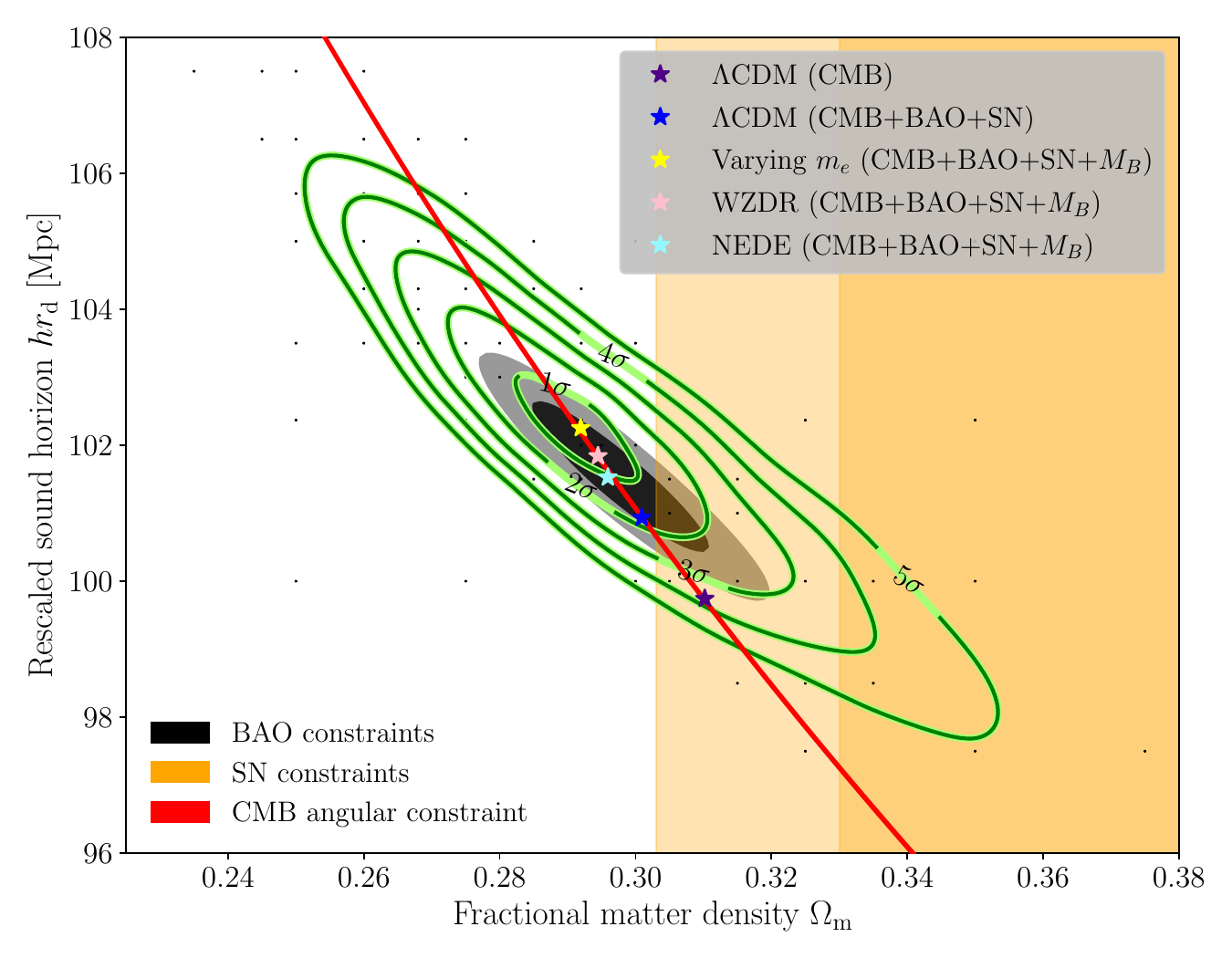}
    \caption{Same as Figure~\ref{fig:rdh_desdovekie} but using Union3 supernovae instead.}
    \label{fig:rdh_union3}
\end{figure}

The main conclusions of this {\it Letter} are robust against the choice of a particular supernova dataset. In Figures~\ref{fig:rdh_pantheonplus} and \ref{fig:rdh_union3} we reproduce Figure~\ref{fig:rdh_desdovekie} for Pantheon+ \cite{Brout:2022vxf} and Union3 \cite{Rubin:2023jdq} supernova datasets, respectively. The \crossing{} does not move by a significant amount in each case, though for the tighter DES Dovekie supernovae it lies closer to the edge of the $2\sigma$ BAO contours. Interestingly, it always remains centered on the CMB angular constraint, despite no CMB data being used in the construction of these iso-contours.

\begin{figure}
    \centering
    \includegraphics[width=1\linewidth]{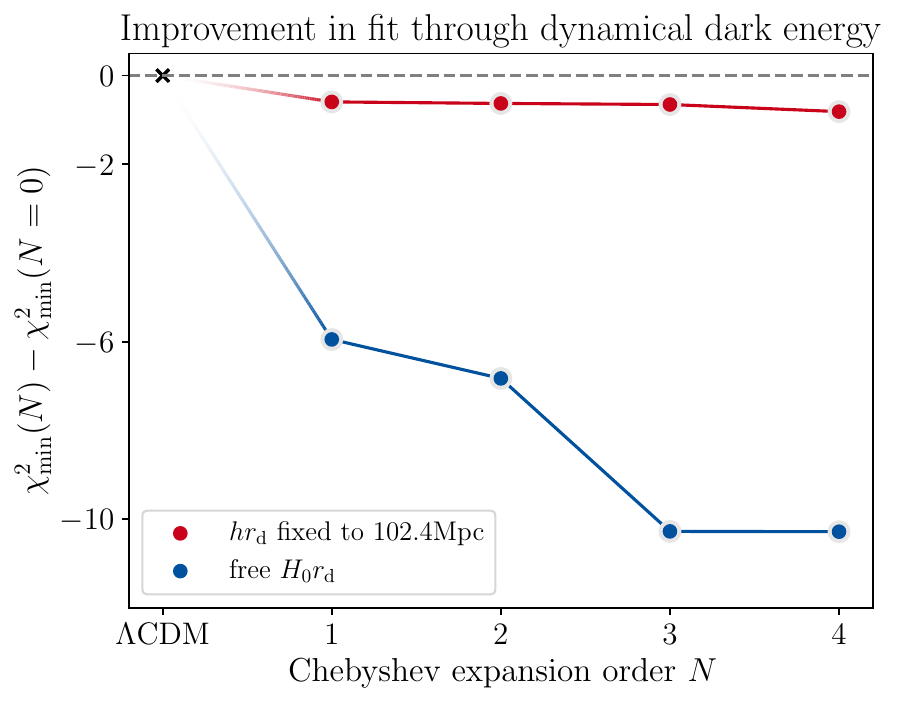}
    \caption{Improvement in best fitting $\chi^2$ compared to a constant dark energy model for a fit to BAO+SN data for various orders of the Chebyshev expansion of the dark energy density. Note that the value at $N=0$ corresponds to a cosmological constant (here marked as $\Lambda$CDM), and since we show differences in $\chi^2$ here, the value there is 0 by definition.}
    \label{fig:chi2}
\end{figure}

In Figure~\ref{fig:chi2} we show the improvement in best-fitting $\chi^2$ from a more flexible dark energy density as a function of redshift as opposed to a cosmological constant (marked there $\Lambda$CDM). When $h r_\mathrm{d}$ is fixed to a high value, there is only a small improvement from allowing more flexible dark energy models, whereas for a free value of $h r_\mathrm{d}$ there is a stronger improvement from allowing more flexible dark energy models.

In the main text we write that we use the Mahalabonis distance to convert $w_0,w_a$ contours from individual fits to SN and BAO data to preference for dynamical dark energy in terms of a number of $\sigma$. Here we expand pedagogically on that statement. The Mahalabonis distance is simply the covariance-weighted distance between two points, computed as
\begin{equation}\label{eq:maha}
    D_\mathrm{Maha}^C(\mu_1, \mu_2) = (\mu_1-\mu_2)^T C^{-1} (\mu_1-\mu_2)~.
\end{equation}
For a proper positive definite covariance matrix, it is easy to see that it provides a proper measure of distance. More importantly for this letter, it can be seen as a fast approximation of the full preference for dynamical dark energy given nearly-Gaussian contours in the $w_0$ and $w_a$ parameters of the CPL parameterization of dark energy. In principle, we could also compute a parameter shift metric as in \cite{Schoneberg:2026vaf}, but in a Gaussian scenario this will reduce naturally to the ordinary Mahalabonis distance between two posteriors, $D_\mathrm{Maha}^{C_1+C_2}(\mu_1,\mu_2)$ with $\mu_i$ and $C_i$ for $i \in \{1,2\}$ being the respective means and covariances. If one of these is a single point (i.e., because we want to know the preference compared to a cosmological constant, which is the single point $w_0=-1, w_a=0$), then the associated covariance contribution vanishes in the limit, and we recover exactly Equation~(\ref{eq:maha}). Given the near-Gaussian approximation, it is also straightforward to show that $D_\mathrm{Maha}^C(\mu_1,\mu_2)$ follows a $\chi^2_d$ distribution with as many degrees of freedom as dimensions $d$ in the mean vector. Therefore, to convert the distance into a significance, we can then simply evaluate
\begin{equation}
    n = \sqrt{2} \mathrm{erf}^{-1}(F_{\chi^2_d}[D_\mathrm{Maha}^C(\mu_1,\mu_2)])
\end{equation}
where $F_{\chi^2_d}$ is the cumulative distribution function of the ${\chi^2_d}$ distribution, $\mathrm{erf}$ is the error function, and $n$ is the equivalent $z$-score (significance in terms of $\sigma$ levels).

For any fixed point in $(\Omega_\mathrm{m}\,, h r_\mathrm{d})$ space,  we can then run a full CPL dark energy fit to BAO and SN data with two free parameters $(w_0\,, w_a)$, infer the $2\times2$ covariance matrix $C$ from the contours (which are typically close to Gaussian), evaluate the corresponding Mahalabonis distance between the mean $(w_0,w_a)$ and $(-1,0)$, and compute its associated significance. Therefore, we can map any point in the ($\Omega_\mathrm{m}$\,,$h r_\mathrm{d}$) space to a significance in terms of $\sigma$ for dynamical dark energy (based on a CPL parameterization) at that point. Then, using a sufficiently dense grid (displayed as the small black dots in Figures.~\ref{fig:rdh_desdovekie}, \ref{fig:rdh_pantheonplus}, and \ref{fig:rdh_union3}, one can easily determine the iso-contours and where the minimal level of significance occurs, defining the position of the \crossing{}.

We could have taken as well the $\Delta_\mathrm{shift}$ metric of \cite{Schoneberg:2026vaf} or used a Frequentist approach, but given the near-Gaussian nature of the contours throughout the parameter space, the results would be equivalent.

\FloatBarrier

\end{document}